\documentstyle[prl,aps,preprint,eqsecnum]{revtex}
\begin{document}
\draft
\input epsf

\title{Einstein--Proca model: spherically symmetric solutions}

\author{Yuri N.\ Obukhov\footnote{Permanent
    address: Department of Theoretical
    Physics, Moscow State University, 117234 Moscow, Russia} and
    Eugen J.\ Vlachynsky}
\address
{Department of Mathematics,
  University of Newcastle, Newcastle, NSW 2308, Australia}

\maketitle

\bigskip
\bigskip

\begin{abstract}
The Proca wave equation describes a classical massive spin 1 particle.
We analyze the gravitational interaction of this vector field. In particular,
the spherically symmetric solutions of the Einstein-Proca coupled system
are obtained numerically. Although at infinity the metric field approaches
the usual Schwarzschild (Reissner-Nordstr\"om) limit, we demonstrate the
absence of black hole type configurations.
\end{abstract}


\section{Introduction}

A massive vector meson (spin 1 particle with a non-trivial mass) is
described by a one-form field which obeys the Proca wave equation
\cite{Proca}. Early development of the Proca theory was concerned with
the classical and quantum electrodynamics of a massive photon. However,
strong experimental limits on photon's mass (see, e.g., \cite{gold}) in
combination with theoretical arguments based on the idea of gauge invariance 
(which ultimately led to the standard model of electroweak interactions) 
have closed the electrodynamical chapter in the history of this theory. 
A further discussion of the differences between the Proca and 
electromagnetic fields can be found in \cite{comay}.

At present, interest in the Proca field is twofold. Firstly, the
Proca model presents a convenient theoretical ``laboratory'' for the
study of Lagrangian and Hamiltonian theories with second class
constraints \cite{toymodel}. Secondly, although it is irrelevant for the
electrodynamics, a massive vector meson often appears in the spectra of 
many non-trivial field theoretical models, including some classes of 
generalized theories of gravity. In connection with this, it is interesting 
to investigate the specific physical effects arising in such models due 
to the interaction of Proca particles with electromagnetic, gravitational 
and other physical and geometrical fields.

The interaction of spin 1 field with electromagnetic field is known to be
free of algebraic inconsistencies as well as of acausal wave propagation
($v>c$) when the coupling is minimal (or modified by the addition of an 
anomalous magnetic dipole moment). However, acausal propagation anomalies 
arise for more general interaction Lagrangians \cite{velo}. Similarly, 
acausal propagation takes place (along with algebraic inconsistencies) 
for a Proca field coupled minimally to external torsion field \cite{acau}.
Different aspects of the interaction of classical and quantum vector 
field with torsion have been analyzed recently in \cite{torsion}.

Early studies of the gravitational interaction of the Proca field
were centered around the black hole issue. Qualitative analysis of
the self-consistent Einstein-Proca system revealed the absence of black
hole type solutions (possessing a regular horizon) with an external vector
meson ``hair'' \cite{beken,isaev,adler}. At the same time several exact
spherically symmetric solutions of the Proca wave equation on the classical
Schwarzschild background spacetime were obtained \cite{frolov,got,linet}.
Assuming that a massive vector field source was located on a thin {\it
spherical shell} outside the Schwarzschild horizon, it was demonstrated in
\cite{frolov} that the meson field may change the structure of the spacetime
near the central singularity. For {\it point} vector field sources
located at the origin \cite{got}, or at a finite distance from the origin
\cite{linet}, it was shown that the range of the meson field is reduced by
the metric gravitational field. The energy-momentum invariant was found to
be divergent on the Schwarzschild horizon. However, it should be noted that,
contrary to the Abelian Proca case, the non-Abelian massive vector field
(with mass of a Yang-Mills field coming from a spontaneous symmetry
breaking mechanism) may form a black hole type configuration \cite{green}.
The results of numerical analysis of the spherically symmetric gravitationally
interacting {\it complex} spin 1 field have been reported recently in
\cite{rosen}. In this case the Einstein-Proca system admits everywhere
regular ``boson star''-type solutions (cf. with massive scalar boson
stars \cite{ruffini,fs}).

A direct motivation for our current study comes from the metric-affine
theory of gravity (MAG). In Einstein's general relativity the spacetime
geometry is described by the curvature 2-form $R^{\alpha\beta}$. In MAG
two post-Riemannian structures are introduced: the 1-form of nonmetricity
$Q_{\alpha\beta}$ and the torsion 2-form $T^\alpha$. For a comprehensive 
review of this theory see \cite{PR}. Already the early investigations 
\cite{pono,TW} of the models with the simplest possible MAG Lagrangians, 
which include only a linear Hilbert term, quadratic segmental curvature 
invariant, and a single trace torsion or Weyl nonmetricity square term, 
have shown that an {\it effective Einstein--Proca theory} arises naturally 
from the vacuum MAG field equations (cf. also \cite{got,buch}). This result 
was subsequently extended to a very general family of MAG Lagrangians 
\cite{magex,tuck,geom}. In all these models the effective Proca field 
describes the triplet of post-Riemannian one-forms which are proportional 
to each other: the Weyl covector $Q :=g^{\alpha\beta}Q_{\alpha\beta}/4$, the 
torsion trace $T:= e_\alpha\rfloor T^\alpha$, and the nonmetricity one-form 
$\Lambda := \vartheta^\alpha e^\beta\rfloor Q_{\alpha\beta} - Q$. The mass 
of the effective vector particle is constructed from the coupling constants 
of the MAG Lagrangian. For a complete review of the known exact solutions 
of MAG see \cite{FA}.

In this paper we study the spherically symmetric static solutions of the 
coupled Einstein-Proca system of field equations. A preliminary analysis of 
the limiting cases of this problem shows a possibility of solutions which
combine the exponential ``Yukawa'' type behavior of the Proca potential
at the origin with the asymptotically Schwarzschild solution far away
from the source. We will present the corresponding solutions which have
been obtained by the application of numerical integration techniques.

Our main conventions and notation are taken from \cite{PR}. In particular,
the $\eta$-basis of the exterior algebra is constructed from a coframe
one-form $\vartheta^\alpha$ with the help of the Hodge duality operator:
$\eta^\alpha = {}^\ast\vartheta^\alpha,\eta^{\alpha\beta} = {}^\ast(
\vartheta^\alpha\wedge\vartheta^\beta), \eta^{\alpha\beta\gamma} = {}^\ast
(\vartheta^\alpha\wedge\vartheta^\beta\wedge\vartheta^\gamma)$. The dual 
frame is denoted as $e_\alpha$. The Greek indices $\alpha,\beta,\dots = 0,
\dots,3$ label anholonomic components, and the metric signature is $(-,+,+,+)$.

\section{Einstein-Proca theory}

The Lagrangian four-form of the Einstein-Proca system reads
\begin{equation}
  V = -\,\frac{1}{2\kappa}\,R^{\alpha\beta}\wedge\eta_{\alpha\beta} -
  \frac{1}{2}\,\left(dA\wedge{}^*dA + m^2\,A\wedge{}^*A\right),\label{lagr}
\end{equation}
where $\kappa$ is the gravitational constant ($\kappa=\ell^2$) and
$m$ is the rest mass of the vector field $A$. The corresponding field
equations arise from the independent variation of the action with respect
to the coframe and the Proca one-forms, and read:
\begin{eqnarray}
  d\,{}^*\!dA + m^2\,{}^*\!A &=& 0,\label{proca1}\\
  \frac{1}{2}\,R^{\beta\gamma}\wedge\eta_{\alpha\beta\gamma} &=&
  \kappa\Sigma_\alpha.\label{ein}
\end{eqnarray}
Here the canonical energy-momentum three-form of the massive vector field
\begin{equation}
  \Sigma_\alpha=\frac{1}{2}\,\left\{(e_{\alpha}\rfloor dA)\wedge{}^{\ast}dA
    - (e_{\alpha}\rfloor{}^{\ast}dA)\wedge dA +
    m^2\left[(e_{\alpha}\rfloor A)^{\ast}\!A
      + (e_{\alpha}\rfloor^{\ast}\!A)\wedge A\right]\right\},
\end{equation}
represents the usual source of the gravitational field.

It is worthwhile to recall the relationship of (\ref{lagr})-(\ref{ein}) to
the MAG theory. As we have mentioned already, the same physical system
arises in MAG as an effective system \cite{pono,TW,magex,tuck,geom,FA} in 
which the effective covector Proca field is (in the notations of our previous
paper \cite{geom})
\begin{equation}
  A=\sqrt{z_4}\,k_0\,\phi.\label{effA}
\end{equation}
Here $\phi$ determines the three nontrivial post-Riemannian pieces of
nonmetricity and torsion (the triplet of one-forms)
\begin{eqnarray}
  &&{}^{(1)}T^{\alpha}={}^{(3)}T^{\alpha}=0,\quad\quad
  {}^{(1)}Q_{\alpha\beta}={}^{(2)}Q_{\alpha\beta}=0,\label{TQ0}\\%
  &&Q=k_0\phi,\quad \Lambda=k_1\phi, \quad T=k_2\phi.\label{triplet}
\end{eqnarray}
The effective mass $m^2$ of the vector particle
and the constants $k_0, k_1, k_2$ 
are constructed from the original coupling constants of the MAG Lagrangian 
which contains all possible quadratic invariants of the torsion and 
nonmetricity (11 terms) together with the linear Hilbert type term
(multiplied by the constant $\kappa$) and the Weyl segmental curvature
quadratic term (multiplied by the constant $z_4$). See \cite{geom} and 
\cite{FA} for more details (note, however, that in the present paper we 
assume that the cosmological constant is zero). 

\section{Spherically symmetric static case}

In terms of the local time and space  coordinates $(\tau, r, \theta, \phi)$,
the general spherically symmetric ansatz for the coframe can be written as
\begin{equation}
  \vartheta^{\hat 0} =\, f\,d\,\tau\,,\quad\vartheta^{\hat 1} =\, {g\over f}
  \, d\, r\,,\quad\vartheta^{\hat 2} =\, r\, d\,\theta\,,\quad
  \vartheta^{\hat 3} =\, r\,\sin\theta \, d\,\phi\,.\label{frame1}
\end{equation}
The geometrical meaning of the function $g(r)$ becomes evident when one
computes the volume four-form
\begin{equation}
  \eta =\vartheta^{\hat 0}\wedge\vartheta^{\hat 1}\wedge\vartheta^{\hat 2}
  \wedge\vartheta^{\hat 3}=g\,r^2\,\sin\theta d\tau\wedge dr\wedge d\theta
  \wedge d\phi.\label{eta}
\end{equation}
Thus, $g(r)$ measures the deviation of $\eta$ from the standard spherically
symmetric spacetime volume form. In a regular oriented spacetime domain we
naturally have to assume
\begin{equation}
  0<g(r)<\infty.\label{reg-g}
\end{equation}

The general static spherically symmetric configuration of the coupled
Einstein-Proca system is described by the three functions $f=f(r), g=g(r)$,
and $u=u(r)$ which enter the spherically symmetric ansatz for the
Proca field as follows
\begin{equation}
  A={u\over rf}\,\vartheta^{\hat 0} = {u\over r}\,d\tau.\label{A1}
\end{equation}
  Substitution of (\ref{frame1})-(\ref{A1}) into the Proca field
equation (\ref{proca1}) results in
\begin{equation}
  \left\{{1\over r^2}\,{f\over g}\,\left[{r^2\over g}\left({u\over r}\right)'
    \right]' - {m^2\,u\over rf}\right\}\,\eta_{\hat 0} =0,
\end{equation}
or, equivalently,
\begin{equation}
  u'' - {g'\over g}\left(u' - {u\over r}\right) -
  {m^2g^2\over f^2}\,u=0.\label{proca2}
\end{equation}

A direct calculation of the energy-momentum 3-form yields
\begin{eqnarray}
  \Sigma_\alpha &=&{1\over 2r^2g^2f^2}\left\{-\,f^2\left(u' - {u\over r}
    \right)^2 + m^2g^2u^2\right\}\delta^{\hat 1}_\alpha\,\eta_{\hat 1}
      \nonumber\\%
  && + {1\over 2r^2g^2f^2}\left\{f^2\left(u' - {u\over r}\right)^2 +
    m^2g^2u^2\right\}\left(-\delta^{\hat 0}_\alpha\,\eta_{\hat 0} +
    \delta^{\hat 2}_\alpha\,\eta_{\hat 2} +
    \delta^{\hat 3}_\alpha\,\eta_{\hat 3}\right).\label{sigma}
\end{eqnarray}

For the sake of completeness we also write down the Einstein 3-form
\begin{eqnarray}
  \frac{1}{2}\,R^{\beta\gamma}\wedge\eta_{\alpha\beta\gamma}&=&
  {f^2\over r^2g^2}\left\{ \left[2r\,{f'\over f} + 1 - {g^2\over f^2}\right]
    \delta^1_\alpha\,\eta_1 + \left[ 2r\left({f'\over f} -
        {g'\over g}\right) + 1 - {g^2\over f^2}\right]
    \delta^{\hat 0}_\alpha\,\eta_{\hat 0}\right\}\nonumber\\%
  && + {f^2\over r^2g^2}\left\{ r^2\,{f''\over f} + \left(r\,{f'\over f}
    \right)^2 - r^2\,{f'\over f} {g'\over g} + 2r\,{f'\over f} - 
    r\,{g'\over g}\right\}\left(\delta^{\hat 2}_\alpha\,\eta_{\hat 2} +
    \delta^{\hat 3}_\alpha\,\eta_{\hat 3}\right).\label{lhs}
\end{eqnarray}

Inserting (\ref{sigma}) and (\ref{lhs}) into (\ref{ein}) we find (after some
algebra) that the equations corresponding to $\alpha={\hat 0}, {\hat 1}$ are
\begin{eqnarray}
  2\left(r\,(f^2)' + f^2 - g^2\right) &=& \kappa\,m^2{g^2\over f^2}u^2
  -\,\kappa\left(u' - {u\over r}\right)^2,\label{f1}\\%
  r\,(g^2)' &=& \kappa\,m^2{g^4\over f^4}u^2.\label{g1}
\end{eqnarray}
Furthermore, it is straightforward to see that the (second order) equation
corresponding to
$\alpha={\hat 2}, {\hat 3}$ is a consequence of (\ref{f1})-(\ref{g1})
and (\ref{proca2}).

The scalar invariant $|\Sigma|:={}^\ast(\Sigma_\alpha\wedge{}^\ast
\Sigma^\alpha)$ characterizes the ``magnitude'' of the energy-momentum of
the massive vector field. Using (\ref{sigma}), we find that
\begin{equation}
  |\Sigma|={}^\ast(\Sigma_\alpha\wedge{}^\ast\Sigma^\alpha)=
  {1\over r^4g^4}\left\{\left(u'-{u\over r}\right)^4
    +  m^2\,u^2\,{g^2\over f^2}\left(u'-{u\over r}\right)^2
    +  m^4\,u^4\,{g^4\over f^4}\right\}.\label{sigma2}
\end{equation}

\section{Preliminary analysis}\label{prelim}

Before we start the study of the complete system it is instructive
to recall two particular cases: namely, massless vector particle in
curved spacetime and massive vector particle in Minkowski spacetime.

For the {\it massless} vector particle
\begin{equation}
 m=0,
\end{equation}
and one finds, from (\ref{g1}), that $g = g_0 = const$. 
Consequently, the vector field equation can now be easily integrated 
to give the usual Coulomb solution
\begin{equation}
  u=q=const\quad \Longrightarrow\quad A = {q\over r}\,d\tau.\label{coul}
\end{equation}
[Strictly speaking, the general solution reads $u=q + \beta r$, but one
can put $\beta=0$ since it contributes only an exact form to $A$].
Turning to (\ref{f1}), we immediately recover the well known
Reissner-Nordstr\"om solution
\begin{equation}
  f^2 = g_0^2\left(1 - {2M\over r} + \kappa {q^2\over 2g_0^2r^2}
\right),\label{reis}
\end{equation}
with the integration constant $M$ interpreted as the mass of the 
gravitating source and $q/g_0$ as its electric charge.

On the other hand, for the Minkowski spacetime the metric functions
are $f=g=1$ and the solution of the Proca field equation (\ref{proca2})
yields the well known Yukawa potential
\begin{equation}
  u=q\,e^{-mr}\quad \Longrightarrow \quad
  A = {q\,e^{-mr}\over r}\,d\tau.\label{yukawa}
\end{equation}
This shows that the vector field is practically zero at distances
much greater than the typical length $r_0=1/m$.

We expect that a spherically symmetric configuration of coupled Einstein 
and Proca fields will combine both of the typical features of the above
limiting cases. Namely, for small values of mass $m$ there will be a large
part of space inside the sphere of radius $r_0=1/m$ where the function
$u$ is to a high degree of approximation constant. In this region the
exact solution will naturally be approximated by (\ref{coul}) and the metric 
will assume the familiar Reissner-Nordstr\"om form (\ref{reis}). However, 
due to its massiveness the field $A$ will remain, also in curved spacetime, 
confined to a finite spatial volume, whereas for $r\rightarrow\infty$ one 
expects a fast decay $u\rightarrow 0$ which leaves one with pure 
Schwarzschild metric.

The following observation will be very useful in the discussion of exact
solutions. Multiplying (\ref{proca2}) by $u/g$ and using the Leibniz rule
one finds that
\begin{equation}
  \left[{u\over g}\left(u'-{u\over r}\right)\right]'=
  {1\over g}\left\{\left(u'-{u\over r}\right)^2 +
    m^2\,u^2\,{g^2\over f^2}\right\},\label{procint}
\end{equation}
where the right-hand side is positive definite in a regular spacetime region.

Identity (\ref{procint}) represents a particular case of the general
relation
\begin{equation}
  db = -(dA\wedge{}^\ast dA + m^2\,A\wedge{}^\ast A),\label{db}
\end{equation}
where $b:=-A\wedge{}^\ast dA$. The latter identity holds true for all
solutions of the Proca field equation (\ref{proca1}).

\section{Dimensionless systems}

The general system is nonlinear and apparently cannot be integrated
analytically. Consequently, we will present the results of numerical
integration in the remainder of this paper. Before we begin the numerical 
analysis, we introduce a new {\it dimensionless} radial variable
\begin{equation}
  \rho:={r\over \sqrt{\kappa}},
\end{equation}
which allows us to rewrite the system (\ref{f1}), (\ref{g1}), and
(\ref{proca2}) in the form
\begin{eqnarray}
  2\left(\rho\,{dF\over d\rho} + F - G\right) &=& K\,{G\over F}\,u^2 -
  \left({du\over d\rho} - {u\over\rho}\right)^2,\label{f2}\\%
  \rho\,{dG\over d\rho} &=& K\,{G^2\over F^2}\,u^2,\label{g2}\\%
  {d^2u\over d\rho^2}&=& K\,{G\over F}\,u + {1\over 2G}\,{dG\over d\rho}
  \left({du\over d\rho} - {u\over\rho}\right).\label{u2}
\end{eqnarray}
Here we have introduced a dimensionless constant
\begin{equation}
  K:=\kappa\,m^2\label{K},
\end{equation}
and defined the functions
\begin{equation}
  F:=f^2,\qquad G:=g^2.
\end{equation}
Evidently, the dimensionless radial coordinate measures distance from the
origin in units of the Planck length ($\kappa=\ell^2$). At the same time,
the parameter $\sqrt{K}$, being the ratio of the Planck length to the 
Compton length of the vector particle, characterizes the size of the domain 
where the influence of the Proca field on the spacetime geometry 
is significant.

One can, alternatively, study a different dimensionless system after
defining the scaled metric functions
\begin{equation}
  \widetilde{F}:=F/K,\qquad \widetilde{G}:=G/K,
\end{equation}
and introducing a new radial coordinate
\begin{equation}
  \xi := \sqrt{K}\rho=m\,r={r\over r_0}.
\end{equation}
The system (\ref{f2})-(\ref{u2}) then reads
\begin{eqnarray}
  2\left(\xi\,{d\widetilde{F}\over d\xi} + \widetilde{F} -
    \widetilde{G}\right) &=& {\widetilde{G}\over \widetilde{F}}\,u^2
   -\left({du\over d\xi} - {u\over\xi}\right)^2,\label{f3}\\%
  \xi\,{d\widetilde{G}\over d\xi} &=& {\widetilde{G}^2\over
    \widetilde{F}^2}\,u^2,\label{g3}\\%
  {d^2u\over d\xi^2}&=& {\widetilde{G}\over \widetilde{F}}\,u +
  {1\over 2\widetilde{G}}\,{d\widetilde{G}\over d\xi}
  \left({du\over d\xi} - {u\over\xi}\right).\label{u3}
\end{eqnarray}
In this form the equations no longer contain a free parameter (such as $K$)
and the new dimensionless coordinate measures distance
in units of the characteristic (``Compton wavelength'') scale $r_0$. It is
convenient to use both dimensionless systems. The advantage of
(\ref{f3})-(\ref{u3}) lies in the absence of $K$, whereas the
equations (\ref{f2})-(\ref{u2}) are more transparent from the physical
point of view when one considers limits of small
and big mass $m$.

\section{Conditions at the origin and at infinity}

Before one can start the numerical integration, an appropriate set of 
initial conditions must be specified. Unfortunately, the solution cannot 
be represented by analytic power series expansion for $u, F, G$ at the 
origin in view of the apparent singularity at $\rho=0$.

Instead, one can verify that for small values of $\rho$, irrespective
of the value of $K$, there is an approximate solution of the form
\begin{eqnarray}
  u&\approx& q + b\,\rho,\label{u0}\\%
  F&\approx& {1\over 2}\,{q^2\over\rho^2},\label{F0}\\%
  G&\approx& {1\over {{1\over c} - {K\over q^2}\rho^4}},\label{G0}
\end{eqnarray}
where $q, b, c$ are parameters which determine the initial
conditions in the neighborhood of the origin $\rho=0$.

At infinity, $\rho\rightarrow\infty$, following the physical discussion
in Sect.~\ref{prelim}, we expect approximate behavior of the form
\begin{eqnarray}
  u&\longrightarrow& u_0\,\exp(-\sqrt{K}\rho),\label{uI}\\%
  F&\longrightarrow& g_0^2\left(1 - {2M\over\rho}\right),\label{FI}\\%
  G&\longrightarrow& g_0^2,\label{GI}
\end{eqnarray}
where $u_0, g_0$ are constants. 
The condition (\ref{uI}) means that the massive vector field is nontrivial
only inside a sphere of a finite radius $1/\sqrt{K}$ (``Yukawa-type''
behavior). On the other hand, the conditions (\ref{FI})-(\ref{GI})
specify purely Schwarzschild asymptotic metric. A more precise form of
the limit (\ref{GI}) is easily obtained after substituting (\ref{uI}) and
(\ref{FI}) into (\ref{f2})-(\ref{u2}):
\begin{equation}
G\approx g_0^2\left(1 + K\,u_0^2\,{\rm Ei}(-2\sqrt{K}\rho)\right),\label{GI1}
\end{equation}
where ${\rm Ei}(x)=\int\limits_{-\infty}^x {\frac {e^t} t}dt$ is the
integral exponential function. It is worthwhile to recall that
asymptotically, for $x\rightarrow\infty$, one has ${\rm Ei}(-x)\approx
-\,e^{-x}/x$.

We will use the asymptotic conditions (\ref{uI})-(\ref{GI}), (\ref{GI1})
in the numerical analysis of the problem under consideration.

\section{Absence of solutions with horizons}\label{nohor}

In this section we show that the spherically symmetric Einstein-Proca
system does not admit asymptotically flat solutions with horizons. 
The absence of black holes for a massive vector field was first 
demonstrated by Bekenstein \cite{beken}.

Let us consider an arbitrary regular solution $u(r)$ which vanishes at
{\it two} points $r_1$ and $r_2>r_1$: $u(r_1)=u(r_2)=0$. Then $u(r)=0$ for
all $r_1\leq r\leq r_2$. Indeed, integrating the identity (\ref{procint})
from $r_1$ to $r_2$, one finds that
\begin{equation}
  \int\limits_{r_1}^{r_2}{1\over g}\left\{\left(u'-{u\over r}\right)^2 +
    m^2\,u^2\,{g^2\over f^2}\right\} dr =
  {u(r_2)\over g(r_2)}\left(u'(r_2)-{u(r_2)\over r_2}\right) -
  {u(r_1)\over g(r_1)}\left(u'(r_1)-{u(r_1)\over r_1}\right)=0.
  \label{int12}
\end{equation}
Since the integrand is positive definite, the vanishing of the integral 
leads to the above conclusion.

Consequently, a nontrivial solution $u(r)$ which vanishes {\it asymptotically} 
at $r_2=\infty$ (thus satisfying the condition (\ref{uI})) cannot have zeros
at any finite $r_1$ (since then the solution would be trivial: $u(r)=0$ for
$r\geq r_1$).

This leads to the absence of the black hole type solutions of the system
(\ref{f2})-(\ref{u2}). In order to see this, let us recall that a black
hole necessarily possesses a horizon. Quite generally, on a spacetime manifold 
$\cal M$ a horizon is defined as a hypersurface $S:=\{x^i\in {\cal M}\, |\,
\sigma(x^i)=0\}$ such that: (i) the normal vector $n_i:=\partial_i\sigma$ 
is null
\begin{equation}
  n_i\,n^i\vert_S =0,\label{horizon}
\end{equation}
and (ii) $S$ is not an essential singularity. The latter means that all the
curvature invariants as well as the volume 4--form $\eta$ are nonsingular
on the horizon. In particular, the regularity of $\eta$ follows from the
condition (\ref{reg-g}) on the function $g$.

For a spherically symmetric gravitational field configuration, horizon
$S$ is evidently a sphere $\sigma=r=r_h$. Normal vector is then $n_i=
\delta_i^1$. Substituting (\ref{frame1}) into (\ref{horizon}), one obtains
\begin{equation}
  {f^2(r_h)\over g^2(r_h)}={F(r_h)\over G(r_h)}=0.
\end{equation}
Since $G(r_h)$ is finite in view of (\ref{reg-g}), we find that $F$ must
vanish on the horizon $S$
\begin{equation}
  F(r_h)=0.\label{zeroF}
\end{equation}
The last equation formally defines the position of a horizon in the general
spherically symmetric spacetime (\ref{frame1}). Now recall the second 
requirement: a hypersurface $S$ must be free of physical singularities 
in order to be a horizon. Clearly, the energy-momentum invariant scalar 
(\ref{sigma2}) is regular at $r=r_h$ if and only if
\begin{equation}
  u(r_h)=0.\label{zerou}
\end{equation}
Furthermore, if (\ref{zerou}) did not hold then the quadratic curvature
invariants (obtained by using the Einstein field equations (\ref{ein}) 
in the definition of $|\Sigma|$) would diverge at $r_h$ because of the 
last term in (\ref{sigma2}) and (\ref{zeroF}).

Now we are in a position to conclude that there are no solutions with a 
horizon and a nontrivial massive vector field. Indeed, assume the contrary 
is true. Then outside a horizon $S$ the function $u$ is necessarily given by 
$u(r)=0,\, r_h\leq r\leq\infty$, because $u$ vanishes at infinity (\ref{uI}) 
and at the horizon (\ref{zerou}). Consequently, outside $S$, the system 
(\ref{f3})-(\ref{g3}) has the usual Schwarzschild solution $G=1, F=1-{r_h/r}$.
Integrating (\ref{f3})-(\ref{u3}) from $r_h$ to $0$ with the initial 
conditions (\ref{zeroF}) and (\ref{zerou}), we find $u(r)=0$ everywhere. 

Bekenstein's original proof \cite{beken} was based on the assumption
that the three-form $b=b^\alpha\eta_\alpha$ defined in (\ref{db}) is
bounded on the horizon. It is easy to see that $b^\alpha={u\over r^2 fg}
\left(u'- {u\over r}\right)\delta^\alpha_{\hat{1}}$ diverges on the
horizon (\ref{zeroF}) unless $u$ vanishes. Thus, for a massive vector
field, the form $b$ is not only bounded but, in fact, trivial on the horizon.

\section{Numerical solutions}

After fixing the value of the parameter $K$  to the
square of the ratio of the Planck length to the Compton wavelength of the
vector field, one can start numerical integration at an arbitrarily small
$\rho$ with the initial conditions defined by (\ref{u0})-(\ref{G0}).
One is free to choose any initial value for the ``boson charge'' function
$u(0)=q$ ($\neq 0$, otherwise $u$ is trivial everywhere). Solutions
with the correct asymptotic behavior (\ref{uI})-(\ref{GI}) exist
only for fixed values of the parameters $b=u'(0), c=G(0)$. Technically,
the numerical integration can start at a point arbitrarily close to the
origin for every chosen values of $K$ and $q$. In order to obtain the
asymptotic behavior (\ref{uI})-(\ref{GI}), a fine tuning of $b$ and $c$ 
is required which can be achieved similarly to the construction of the 
Bartnik-McKinnon solutions \cite{bart} or of the Abrikosov-Nielsen-Olesen 
vortices (see, e.g. \cite{vort} and references therein). Alternatively, one 
can start the numerical integration at a sufficiently large radius with the 
initial conditions taken from (\ref{uI})-(\ref{GI}) for arbitrary values of
$K, M, u_0$. As a cross-check, we have used both integration schemes. The 
resulting approximate solutions turned out to be completely consistent 
with each other.

\begin{figure}
\centering
\leavevmode\epsfysize=8cm \epsfbox{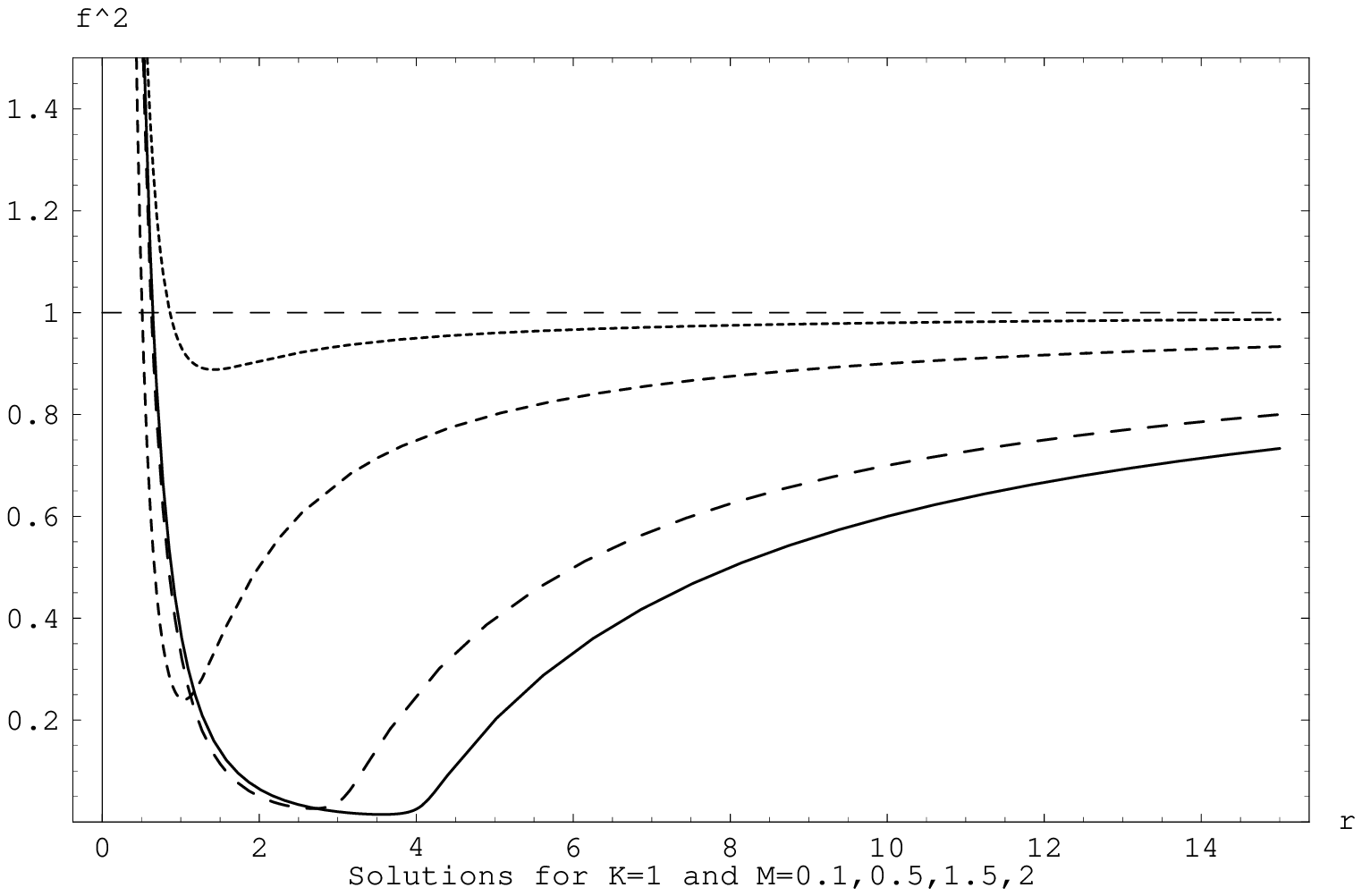}\\
\caption[]
{The metric function $F=f^2$: $K=1$ solutions for the values  $M=0.1$
(dotted line), $M=0.5$ and $M=1.5$ (broken lines), and $M=2$ (solid line).}
\label{figF}
\end{figure}

\begin{figure}
\centering
\leavevmode\epsfysize=8cm \epsfbox{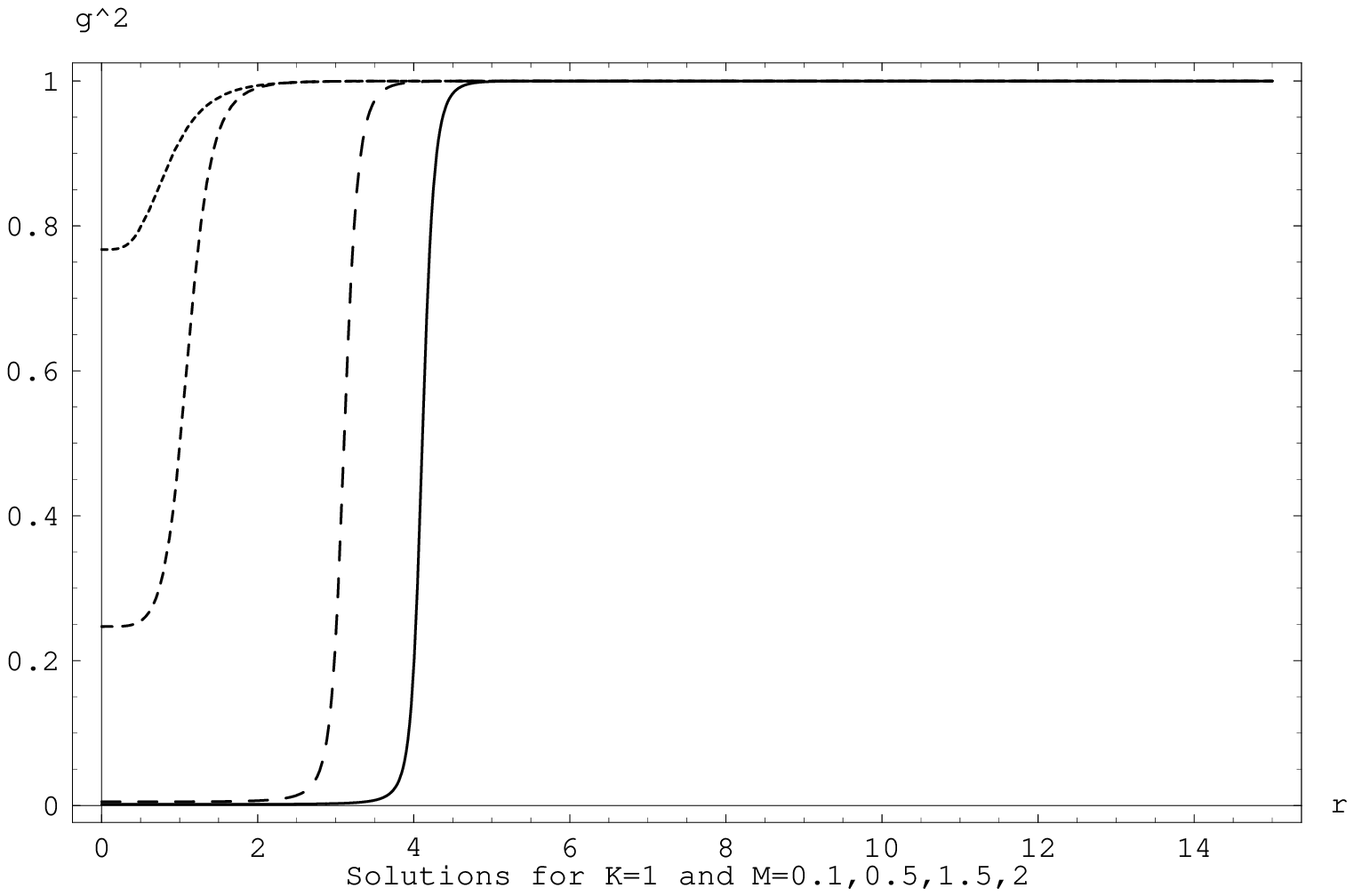}\\
\caption[]
{The metric function $G=g^2$: $K=1$ solutions for the values  $M=0.1$
(dotted line), $M=0.5$ and $M=1.5$ (broken lines), and $M=2$ (solid line).}
\label{figG}
\end{figure}

\begin{figure}
\centering
\leavevmode\epsfysize=8cm \epsfbox{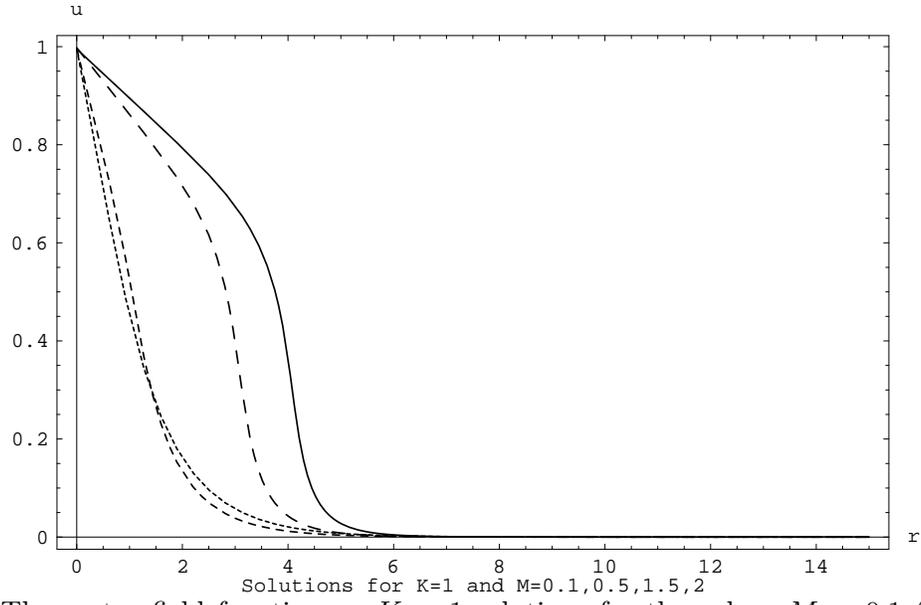}\\
\caption[]
{The vector field function $u$: $K=1$ solutions for the values  $M=0.1$
(dotted line), $M=0.5$ and $M=1.5$ (broken lines), and $M=2$ (solid line).}
\label{figU}
\end{figure}

Particular solutions for various values of $K$, $q$ and $M$ are described in
Tables \ref{table1}-\ref{table3}. The graphical form of the solutions
is presented in Figures \ref{figF},\ref{figG}, and \ref{figU}. In these
figures, the numerical solutions are depicted for $K=1$ and $M=0.1$
(dotted lines), $M=0.5$ and $M=1.5$ (broken lines), and $M=2$ (solid lines).
In all cases we put $g_0 = 1$ which is always possible to achieve by the
redefinition of the time coordinate $\tau\rightarrow g_0\tau$. 

As one can see, the relation between $K$ (formal rest mass of the vector
field) and $M$ (asymptotic total mass of the solution) plays a decisive
role. At the same time, the value of the boson charge $q$ at the origin
is also important.

In agreement with the results of Bekenstein et al \cite{beken,isaev,adler}
and with the preliminary analysis of Sect.~\ref{nohor}, all the numerical
solutions obtained by us are without horizons. They possess a true physical
singularity at the origin which provides us with an example of a {\it naked}
spacetime singularity. Stability of these solutions against small
perturbations will be studied separately.

Recalling that the effective Proca field emerges naturally in the general
metric-affine models, we thus conclude that the presence of the
post-Riemannian geometric objects prevents, in general, a formation of
a black hole in MAG theory. Only in the special case when the MAG coupling 
constants are such that the effective mass vanishes, $m^2 = 0$, the black
holes can be formed \cite{TW,magex}.

\acknowledgments

This work was supported by a grant from the University of Newcastle which
made it possible for YNO to visit Australia. We are grateful to Robin Tucker 
and Tekin Dereli for useful comments, and to Marc Toussaint for a
discussion of the results obtained in this paper.


\bigskip


\begin{table}
  \caption[]{Solutions with fixed values of $q$ and $M$}
  \begin{tabular}{ccc|ccc}
 & $q=4$ and $M=0.5$ & & & $q=2$ and $M=1$ & \\
\tableline
    $K$ & $b$ & $c$ & $K$ & $b$ & $c$\\  %
    \tableline
    0.01  & $-$0.29605 & 0.90444 & $10^{-6}$  & $-$0.00202 & 0.99996\\ %
    0.10  & $-$0.66479 & 0.69257 & 0.01  & $-$0.16267 & 0.84124 \\  %
    1.00  & $-$1.20050 & 0.34457 & 1.00  & $-$0.56351 & 0.08738 \\ %
    10.00 & $-$1.67696 & 0.08873 & 10.00 & $-$0.69666 & 0.01100 \\ %
  \end{tabular}
  \label{table1}
\end{table}

\begin{table}
  \caption[]{$K=1.00$: solutions for $q=1$ with different masses $M$}
  \begin{tabular}{lll}
    $M$ & $b$ & $c$\\ %
    \tableline
    0.10  & $-$0.56959 & 0.76748 \\ %
    0.50  & $-$0.43469 & 0.24706 \\ %
    1.50  & $-$0.13588 & 0.00528 \\ %
    2.00  & $-$0.10007 & 0.00170 \\ %
  \end{tabular}
  \label{table2}
\end{table}

\begin{table}
  \caption[]{Solutions for $M=1$ and different $q$}
  \begin{tabular}{l|cc|cc}
    &\multicolumn{2}{c|}{$q=0.50$}&\multicolumn{2}{c}{$q=1.00$}\\ %
    \tableline
    $K$ & $b$ & $c$ & $b$ & $c$ \\ %
    \tableline
    0.01  & 0.60874 & 0.08438 & 0.25196 & 0.34048 \\ %
    1.00  & $-$0.01728 & 0.00579 & $-$0.21296 & 0.02528 \\ %
    10.00 & $-$0.13175 & 0.00071 & $-$0.33056 & 0.00308 \\ %
  \end{tabular}
  \label{table3}
\end{table}


\begin{references}                    
\bibitem{Proca}
W. Pauli, 
{\sl Rev. Mod. Phys.} {\bf 13} (1941) 203; 
E.M. Corson, {\it Introduction to tensors, spinors, and relativistic wave
equations} (Blackie: London, 1953);
H. Umezawa, {\it Quantum field theory} (North Holland: Amsterdam, 1956).
\bibitem{gold}
A.S. Goldhaber and M.M. Nieto,
{\sl Rev. Mod. Phys.} {\bf 43} (1971) 277.
\bibitem{comay}
E. Comay, 
{\sl Nuovo Cim.} {\bf 113B} (1998) 733; 
P. Hillion and S. Quinnez, 
{\sl Int. J. Theor. Phys.} {\bf 25} (1986) 727. 
\bibitem{toymodel}
N. Banerjee and R. Banerjee,
{\sl Mod. Phys. Lett.} {\bf A11} (1996) 1919; 
J. Camacaro, R. Gaitan, and L. Leal,
{\sl Mod. Phys. Lett.} {\bf A12} (1997) 3081; 
V. Aldaya, M. Calixto, and M. Navarro,
{\sl Int. J. Mod. Phys.} {\bf A12} (1997) 3609; 
Y.-W. Kim, M.-I. Park, Y.-J. Park, and S.J. Yoon, 
{\sl Int. J. Mod. Phys.} {\bf A12} (1997) 4217; 
A.S. Vytheeswaran, 
{\sl Int. J. Mod. Phys.} {\bf A13} (1998) 765; 
M.-I. Park, and Y.-J. Park,
{\sl Int. J. Mod. Phys.} {\bf A13} (1998) 2179. 
\bibitem{velo}
G. Velo and D. Zwanziger, 
{\sl Phys. Rev.} {\bf 188} (1969) 2218. 
\bibitem{acau}
Yu.N. Obukhov, 
{\sl J. Phys.} {\bf A16} (1983) 3795; 
Yu.N. Obukhov, 
in: {\sl Contrib. Papers of 10th Int. Conf. ``General Relativity and
Gravitation'' (Padova, 4-9 July 1983)}, Eds. B. Bertotti, F. De Felice, and
A. Pascolini (Cons. Naz. delle Ric.: Rome, 1983), vol. 1, p. 599-601.
\bibitem{torsion}
M. Seitz, 
{\sl Class. Quantum Grav.} {\bf 3} (1986) 1265; 
R. Spinosa, 
{\sl Class. Quantum Grav.} {\bf 4} (1987) 473; 
V.G. Bagrov, A.Yu. Trifonov, and A.A. Evseevich,
{\sl Class. Quantum Grav.} {\bf 9} (1992) 533. 
\bibitem{beken}
J. Bekenstein, 
{\sl Phys. Rev. Lett.} {\bf 28} (1972) 452; 
J. Bekenstein,  
{\sl Phys. Rev.} {\bf D5} (1972) 1239; 
J. Bekenstein, 
{\sl Phys. Rev.} {\bf D5} (1972) 2403. 
\bibitem{isaev}
G.V. Isaev, {\it Massive vector field and horizon surface},
in: {\sl ``Problems of gravity and elementary particle theory'' / Ed.
K.P. Stanyukovich} (Atomizdat: Moscow, 1976) v. {\bf 7}, p. 138-141
(in Russian).
\bibitem{adler}
S.L. Adler and R.B. Pearson,
{\sl Phys. Rev.} {\bf D18} (1978) 2798. 
\bibitem{frolov}
V.P. Frolov, 
{\sl Gen. Relat. Grav.} {\bf 9} (1978) 569. 
\bibitem{got}
D. Gottlieb, R. Hojman, L.H. Rodrigues, and N. Zamorano,
{\sl Nuovo Cim.} {\bf B80} (1984) 62. 
\bibitem{linet}
B. Leaute and B. Linet, 
{\sl Gen. Rel. Grav.} {\bf 17} (1985) 783. 
\bibitem{green}
B.R. Greene, S.D. Mathur, and C.M. O'Neill, 
{\sl Phys. Rev.} {\bf D47} (1993) 2242; 
M.E. Ortiz, 
{\sl Phys. Rev.} {\bf D45} (1992) R2586; 
K. Lee, V.P. Nair, and E.J. Weinberg,
{\sl Phys. Rev.} {\bf D45} (1992) 2751. 
\bibitem{rosen}
N. Rosen, 
{\sl Found. Phys.} {\bf 24} (1994) 1689. 
\bibitem{ruffini}
R. Ruffini and S. Bonazzola, 
{\sl Phys. Rev.} {\bf 187} (1969) 1767. 
\bibitem{fs}
E.W.~Mielke and F.E.~Schunck, {\it Boson Stars: Early history and recent
prospects}, in: {\sl Proc. of 8th Marcel-Grossmann Meeting on ``Recent
Developments in Theoretical and Experimental General Relativity, Gravitation
and Relativistic Field Theories'', Jerusalem, Israel, 22-27 Jun 1997},
Ed. T.~Piran (World Scientific, Singapore 1999, to appear); e-Print Archive:
gr-qc/9801063.
\bibitem{PR}
F.W. Hehl, J.D. McCrea, E.W.  Mielke, and Y. Ne'eman,
{\sl Phys. Rep.} {\bf 258} (1995) 1.
\bibitem{pono}
V.N. Ponomariov and Yu. Obukhov,
{\sl Gen. Relat. Grav.} {\bf 14} (1982) 309. 
\bibitem{TW}
R.W. Tucker and C. Wang,
{\sl Class. Quantum Grav.} {\bf 12} (1995) 2587. 
\bibitem{buch}
H.A. Buchdahl, 
{\sl J. Phys. A: Math. and Gen.} {\bf A12} (1979) 1235. 
\bibitem{magex}
Yu.N. Obukhov, E.J. Vlachynsky, W. Esser, R. Tresguerres, and F.W. Hehl,
{\sl Phys. Lett.} {\bf A220} (1996) 1; 
E.J. Vlachynsky, R. Tresguerres, Yu.N. Obukhov, and F.W. Hehl,
{\sl Class. Quantum Grav.} {\bf 13} (1996) 3253. 
\bibitem{tuck}
T. Dereli, M. \"Onder, J. Schray, R.W. Tucker, and C. Wang,
{\sl Class. Quantum Grav.} {\bf 13} (1996) L103; 
R.W. Tucker and C. Wang, 
in: {\sl ``Mathematics of gravitation. Part II. Gravitational wave detection",
Proc. of the Workshop ``Mathematical Aspects of Theories of Gravitation'',
(Warsaw, Poland, 26 Feb - 30 Mar 1996)}, Banach Center Publications,
Vol. {\bf 41} (Inst. of Math., Polish Acad. Sci.: Warszawa, 1997) 263. 
\bibitem{geom}
Yu.N. Obukhov, E.J. Vlachynsky, W. Esser and F.W. Hehl,
{\sl Phys. Rev.} {\bf D56} (1997) 7769. 
\bibitem{FA}
F.W. Hehl and A. Mac\'{\i}as, {\it Metric affine gauge theory of gravity.
2. Exact solutions}, {\sl Preprint UAM-I-9902026 (Mexico City Univ.,
Iztapalapa: February 1999)} 25pp; e-Print Archive: gr-qc/9902076; to be
publ. in {\sl Int. J. Mod. Phys.} {\bf A} (1999).
\bibitem{bart}
R. Bartnik and J. McKinnon,
{\sl Phys. Rev. Lett.} {\bf 61} (1988) 141. 
\bibitem{vort}
Yu.N. Obukhov and F.S. Schunck,
{\sl Phys. Rev.} {\bf D55} (1997) 2307. 
\end{references}
\end{document}